# Observation of Orbit-Orbit Torques: Highly Efficient Torques on Orbital Moments Induced by Orbital Currents

Hongyu Chen[1,2], Han Yan[1,2], Xiaorong Zhou[1,2], Xiaoning Wang[1,2], Ziang Meng[1,2], Li Liu[1,2], Guojian Zhao[1,2], Zhiyuan Duan[1,2], Sixu Jiang[1,2], Jingyu Li[1,2], Xiaoyang Tan[1,2], Peixin Qin[1,2]*, Zhiqi Liu[1,2]*

[1]School of Materials Science and Engineering, Beihang University; Beijing 100191, China.

[2]State Key Laboratory of Tropic Ocean Engineering Materials and Materials Evaluation, Beihang University; Beijing 100191, China

*Corresponding authors. Emails: qinpeixin@buaa.edu.cn; zhiqi@buaa.edu.cn

**Abstract**

We study the current-induced torques in bilayers composed of a light 3$d$ metal, chromium, and a rare-earth ferromagnet with finite orbital moments, terbium, utilizing second-harmonic Hall-response measurements. The dampinglike torque efficiency $\xi_{\mathrm{DL}}^{j}$ of chromium is found to be positive and reaches ~3.66 in this system, in sharp contrast to the negative and subtle $\xi_{\mathrm{DL}}^{j}$ in general Cr/ferromagnet heterostructures with quenched orbital moment. We suggest that the orbital currents generated by the orbital Hall effect in Cr can be injected into Tb with negligible loss at the interface and then efficiently interact with the orbital moments. We term such an exotic effect as the orbit-orbit torque (OOT). Our work implies that orbital currents could be harnessed to manipulate the orbital magnetization of materials, which would advance the development of orbitronics.

# 1. Introduction

Spin currents, *i.e.*, the flow of spin angular momenta $S$, have been demonstrated to exert sizable torques on the (spin) magnetic moments of magnets [1,2], which constitutes the building blocks of modern spintronic devices. Heavy metals, such as Pt [3] and Ta [4], are typically employed as spin-current sources in such components, since the generation of spin currents typically requires strong spin-orbit interactions (SOIs). Recently, the flow of the orbital degree of freedom of electrons, *i.e.*, orbital currents, has been proposed as a possible alternative to spin currents for manipulating magnetization, which provokes the emergence of a brand-new field termed orbitronics [5].

The production of orbital currents has been shown irrelevant with SOIs [5–10], so that light materials, such as Ti [11] and Cr [12], can serve as orbital-current generators. Indeed, the spin Hall and spin Rashba-Edelstein effect, wherein charge currents are converted to transverse spin currents, can be regarded as secondary effects of their orbital counterparts under finite SOIs [5–10]. Moreover, giant orbital Hall conductivity on the order of $10^3$ ($\hbar/e$) ($\Omega$ cm)$^{-1}$ have been predicted in several $3d$ metals [13], which is rather appealing for orbitronic applications. Nevertheless, orbital currents seem unable to directly interact with the (spin) magnetic moments of general magnets. In order to utilize the gigantic orbital currents produced by the orbital Hall or orbital Rashba-Edelstein effect to control magnetization, SOIs that can transform orbital angular momenta $L$ back to $S$ shall be resorted to [14]. Experimentally, such *L-S* conversions can be realized in ferromagnets with strong SOIs themselves [15,16], or via a heavy-metal interlayer between the orbital-current source and the ferromagnet [17,18]. The direction of the converted $S$ (parallel or antiparallel to $L$) depends on the sign of the SOIs in the media materials. Consequently, unavoidable loss in the overall torque efficiency could arise at interfaces or due to the strong SOIs in the *L-S* conversion processes.

We note that the aforementioned studies mostly focused on the orbital torques on $3d$ ferromagnets with large spin moments but quenched orbital moment, while whether and how orbital currents can interact with orbital moments remain elusive. Here we employ Cr/Tb bilayers to investigate the interaction between orbital currents and orbital moments. As illustrated in Fig. 1, Cr has been predicted and experimentally demonstrated to exhibit a small negative spin Hall conductivity $\sigma_{SH}^{Cr}$ [16,18–20] but a giant positive orbital Hall conductivity $\sigma_{OH}^{Cr}$ [12,13,16,18]. On the other hand, Tb is a special rare-earth ferromagnet with comparable and parallel spin and orbital moments [21,22] and a negative SOI. As a result, the polarization direction of the orbital currents and the (converted) spin currents generated from Cr shall be opposite. Given that orbital moments are unable to efficiently act on spin moments, a positive overall torque from Cr in such bilayers can unambiguously indicate the interaction between orbital currents and orbital moments.

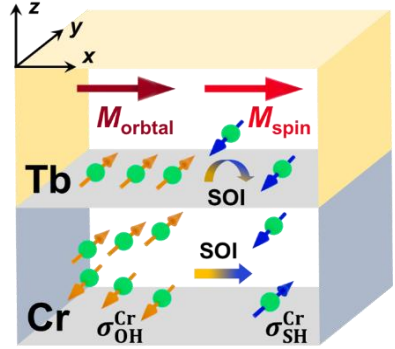

FIG 1. Schematic on the orbital and spin currents in Cr/Tb bilayers. With the application of an electric current along –$x$, the positive orbital Hall conductivity $\sigma_{OH}^{Cr}$ of Cr can generate large transverse orbital currents (orange arrows), which then lead to finite spin currents (blue arrows) with opposite propagation direction owing to the weak negative spin-orbit interaction (SOI) and the resultant small spin Hall conductivity $\sigma_{SH}^{Cr}$ of Cr. When injected to the adjacent Tb layer, the orbital currents can be converted to spin currents with opposite polarization direction under the assistance of the negative SOI in Tb, and thus add up to the injected spin currents to exert torques on the parallel spin and (or) orbital magnetization $M_{spin} + M_{orbital}$. Here we explore the possible interaction between the orbital currents and $M_{orbital}$, which would result in torques opposite to those from the (converted) spin currents.

## 2. Results and Discussions

All the samples in this study were deposited via a magnetron sputtering system with a base pressure of ~7.5 × 10$^{-9}$ Torr on (001)-oriented MgO substrates. The 10-nm-thick Cr layer was grown at 500°C, while the 10-nm-thick Tb layer was deposited at room temperature. An Al layer of ~2 nm was *in situ* capped on top of the Cr/Tb bilayers at room temperature, which would be naturally oxidized upon exposure to atmosphere. The as-deposited films were processed into Hall bars via standard ultra-violet lithography and Ar$^+$ etching. The width of the current channel was 5 μm and the separation of the voltage probes was 25 μm.

The magnetization loops of Cr/Tb at various temperatures are shown in Fig. 2(a). Owing to the relatively low Curie temperature $T_C$ ~220 K of Tb [23], the saturation magnetization $M_S$ decreases from ~858.61 kA m$^{-1}$ at 100 K to ~298.23 kA m$^{-1}$ at 200 K. Note that the $M_S$ ~676.22 kA m$^{-1}$ at 150 K is in good accordance with that of sputtered Tb thin films reported previously [24]. Despite the well-known spin-density-wave antiferromagentism of Cr [25], no prominent exchange bias effect was detected in the Cr/Tb bilayers in our measurements. The extracted exchange-bias field is lower than 10 Oe at 50 K, and decreases to almost zero at 100 K. This can originate from the fact that the Tb layer was deposited at room temperature (above its $T_C$), and that the sample was zero-field cooled to 50 K to perform the magnetic characterizations, so that no effective exchange anisotropy was induced in the sample. The anomalous Hall effect extracted from the Hall measurements of the Hall-bar sample is displayed in Fig. 2(b). It was found that a magnetic field of 3 T cannot completely saturate the anomalous Hall resistance $R_{AHE}$ at 50 K. As a result, only the data collected above 50 K are shown. Owing to the much higher conductivity of the Cr layer compared to that of the Tb layer, the saturation $R_{AHE}$ is relatively small in Cr/Tb, for example, ~10.56 mΩ at 150 K. In addition, the polarity of the anomalous Hall effect in Cr/Tb is opposite to that in Cr/Co [26], which could be a common feature for rare-earth ferromagnets [16,27].

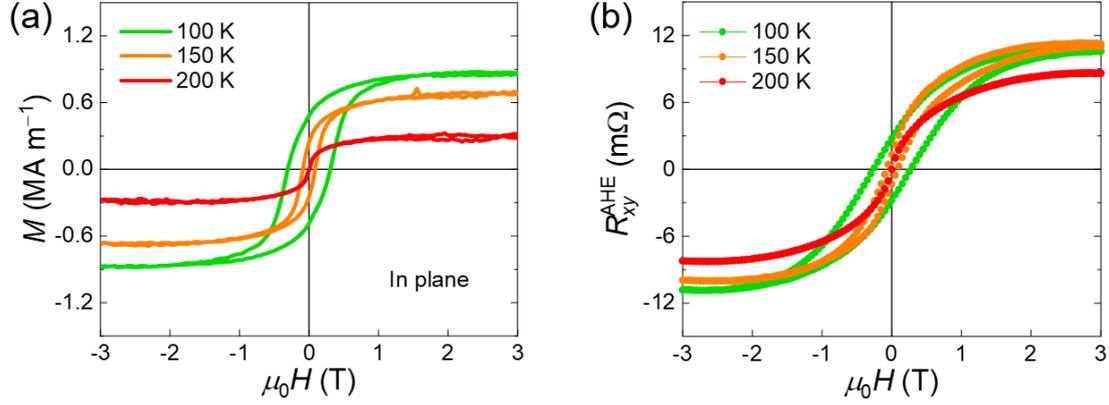

FIG 2. (a) In plane magnetization $M$ and (b) Hall resistance after subtracting the ordinary Hall effect $R_{xy}^{\mathrm{AHE}}$ as a function of applied magnetic fields ($\mu_0 H$) of Cr/Tb at various temperatures.

Based on the above analyses, we selected 150 K, where the exchange-bias effect is negligible and a well-defined saturation $R_{\mathrm{AHE}}$ can be obtained, as the appropriate temperature to study the current-induced torques in Cr/Tb. The angle-scan second-harmonic Hall-response technique [28], in which the first and second harmonic Hall voltages are simultaneously collected while rotating the sample under in-plane static magnetic fields [Fig. 3(a)], was employed to investigate the torques. This is because Cr/Tb does not exhibit a well-defined magnetic easy axis or easy plane but can be magnetically saturated with in-plane fields above 2 T at 150 K (Fig. 2), so that current-induced torques can be treated as perturbations when the applied field is higher than 2 T, and can therefore be separated from thermal effects. We harnessed a Keithley 6221 source meter to apply ac currents of 13 Hz to the sample with a current density of ~$10^7$ A cm$^{-2}$. Two Standford SR830 lock-in amplifiers were employed to measure the first and second harmonic Hall voltages in static magnetic fields ranging from 2 to 3 T.

The first harmonic Hall responses of the sample can be well fitted with a sin2$\varphi$ function [Fig. 3(b)], where $\varphi$ denotes the angle between the applied current and the external magnetic field. This indicates that the out-of-plane projection of magnetization is negligible and is consistent with the saturated in-plane magnetization of the sample. A planar Hall resistance of ~51.88 m$\Omega$ can be deduced from the fitting. The $\varphi$-dependent second harmonic Hall resistances are depicted in Fig. 3(c), which can be well fitted by taking into consideration the contributions from the fieldlike and dampinglike current-induced torques due to orbital/spin currents polarized along y, thermal effects, including the anomalous Nernst effect and the planar Nernst effect, and current-induced transverse Oersted fields. Owing to the high crystalline symmetry of bcc Cr, we do not expect spin or orbital currents with unconventional polarization directions to exist in our samples. It was found that the effective field of the dampinglike torques is positive and giant, reaching ~178.3 Oe [Fig. 3(d)], while the fieldlike torque is negligible [26]. The dampinglike torque efficiency per unit current density $\xi_{DL}^j$ was calculated to be as high as ~3.66 [26], thirty times larger than that of Pt [3,29,30].

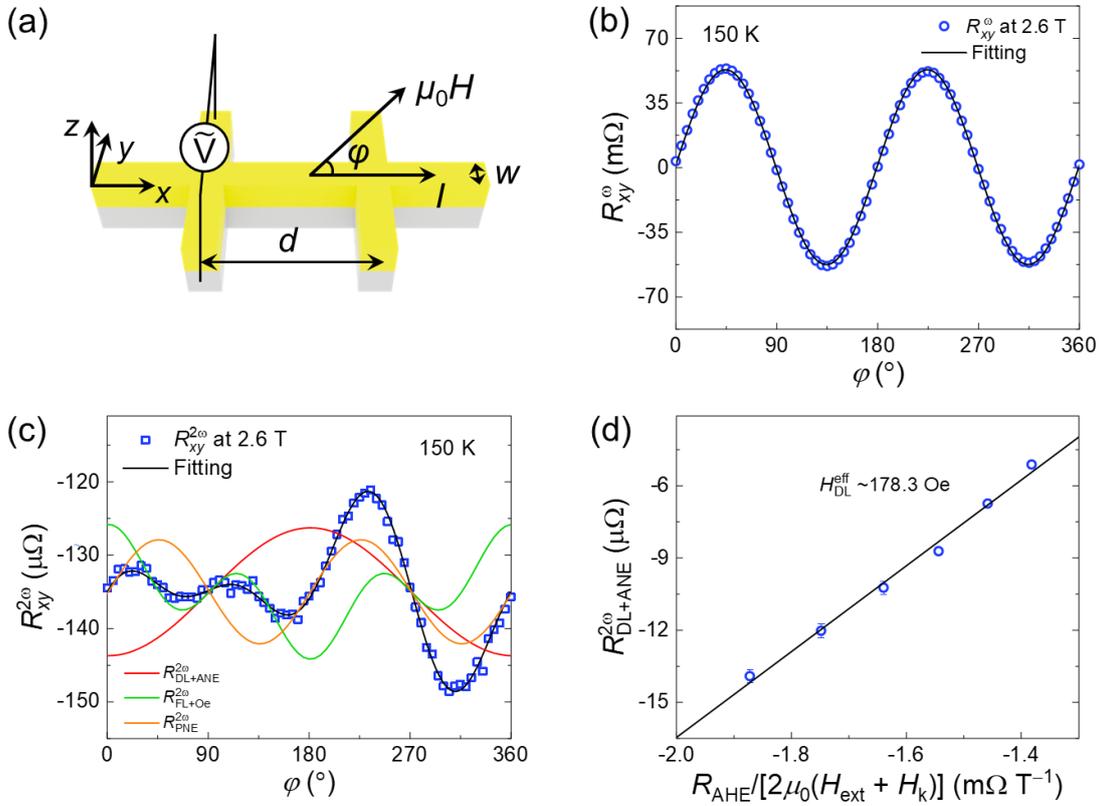

FIG 3. (a) Schematic of the measurement geometry in the second-harmonic Hall-response experiments. The current $I$ is applied along $x$ and the sample is rotated in the $xy$ plane under different in-plane fields. The ratio $w/d$ of the Hall bar is 1/5. (b) First-harmonic Hall resistance $R_{xy}^\omega$ and (c) second-harmonic Hall resistance $R_{xy}^{2\omega}$ as a function of the angle $\varphi$ between $I$ and $\mu_0 H$ in (a) at 150 K. The black lines are fitting curves according to [26]. The contributions to $R_{xy}^{2\omega}$ from the dampinglike torques and the anomalous Nernst effect $R_{DL+ANE}^{2\omega}$, the fieldlike torques and current-induced transverse Oersted fields $R_{FL+Oe}^{2\omega}$, and the planar Nernst effect $R_{PNE}^{2\omega}$ are also shown. (d)

Linear fitting of $R_{\mathrm{DL+ANE}}^{2\omega}$ with respect to $R_{\mathrm{AHE}}/[2\mu_0(H_{\mathrm{ext}} + H_{\mathrm{k}})]$, which yields an effective dampinglike field $H_{\mathrm{DL}}^{\mathrm{eff}}$ of ~178.3 Oe. $R_{\mathrm{AHE}}$, $\mu_0 H_{\mathrm{ext}}$, and $\mu_0 H_{\mathrm{k}}$ denote the anomalous Hall resistance of the sample, the applied static magnetic field, and the anisotropy field of the sample, respectively.

Such a giant $\xi_{\mathrm{DL}}^{j}$ can be generally achieved in topologically nontrivial materials [31,32], and is therefore unusual for Cr/Tb layers. It seems that strong orbital/spin currents are generated by the applied electric currents in this system, which then efficiently exert torques on the magnetization of Tb. Firstly, interfacial effects are not expected to result in such a large $\xi_{\mathrm{DL}}^{j}$ considering the relatively thick thickness (10 nm) of the Cr and Tb layers in our sample. In addition, the self-torques of Tb can be excluded as well, since the Tb layer deposited at room temperature shall be short-circuited by the highly conductive Cr layer sputtered at 500 °C.

We note that Cr has been reported to exhibit a small and negative $\sigma_{\mathrm{SH}}^{\mathrm{Cr}}$, which typically yield a negative $\xi_{\mathrm{DL}}^{j}$ on the order of 0.01 in general Cr/ferromagnet bilayers [16,18,26]. On the other hand, $\sigma_{\mathrm{OH}}^{\mathrm{Cr}}$ has been predicted to be several times larger than the spin Hall conductivity of Pt [13], and has been demonstrated to be positive in sign [12,16,18]. Although the orbital current produced by Cr may induce sizable orbital torques through the strong SOI in Tb, the resultant $\xi_{\mathrm{DL}}^{j}$ shall be negative in sign owing to the negative SOI of Tb [18]. Therefore, we deduce that the measured $\xi_{\mathrm{DL}}^{j}$ can result from an interaction between the orbital currents in Cr and the magnetization of Tb. It is well known that Tb possesses comparable and parallel spin and orbital magnetization [21,22]. Our control experiments on a Cr/Co sample with zero orbital moment demonstrate that the giant orbital current in Cr cannot effectively interact with the spin magnetization [26]. As a result, the observed $\xi_{\mathrm{DL}}^{j}$ in Cr/Tb shall stem from the interaction between orbital currents and orbital magnetization, thus experimentally demonstrating the OOT. We suggest that the transmission of orbital currents and their torques on orbital moments could be immune to strong SOIs, unlike their spin counterparts, and thus results in the observed large $\xi_{\mathrm{DL}}^{j}$. The detailed mechanisms deserve further studies.

## 3. Conclusion

In conclusion, our experiments on the current-induced torques in Cr/Tb layers indicate large OOTs, *i.e.*, remarkable torques of orbital currents on orbital moments. Our work suggests that OOTs could be rather efficient to manipulate the magnetization of ferromagnets with finite orbital moments.

**Acknowledgements**

P.Q. acknowledges the financial support from the National Natural Science Foundation of China (no. 52401300). Z.M. acknowledges the financial support from the National Natural Science Foundation of China (no. 524B2003). L.L. acknowledges the financial support from the National Natural Science Foundation of China (no. 525B2008). Z.L. acknowledges financial supports from the National Natural Science Foundation of China (nos. 52425106, 52121001 and 52271235), the National Key R&D Program of China (no. 2022YFA1602700) and the Beijing Natural Science Foundation (no. JQ23005). This work is supported by National Natural Science Foundation of China (no. U25A20244). This work is supported by the Fundamental Research Funds for the Central Universities. The authors acknowledge the Analysis & Testing Center of Beihang University for the assistance.

# Supplemental Material for

# Observation of Orbit-Orbit Torques: Highly Efficient Torques on Orbital Moments Induced by Orbital Currents

Hongyu Chen[1,2], Han Yan[1,2], Xiaorong Zhou[1,2], Xiaoning Wang[1,2], Ziang Meng[1,2], Li Liu[1,2], Guojian Zhao[1,2], Zhiyuan Duan[1,2], Sixu Jiang[1,2], Jingyu Li[1,2], Xiaoyang Tan[1,2], Peixin Qin[1,2]\*, Zhiqi Liu[1,2]\*

[1]School of Materials Science and Engineering, Beihang University; Beijing 100191, China.

[2]State Key Laboratory of Tropic Ocean Engineering Materials and Materials Evaluation, Beihang University; Beijing 100191, China

\*Corresponding authors. Emails: qinpeixin@buaa.edu.cn; zhiqi@buaa.edu.cn

# 1 Torque analyses via second-harmonic Hall-response technique

In angle-scan second-harmonic Hall-response measurements, the sample is rotated in different static in-plane magnetic fields that are higher than the saturation field. For spin/orbital polarization along $y$, the detected second-harmonic Hall resistance $R_{xy}^{2\omega}$ as a function of the angle $\varphi$ between the applied magnetic field and the current flow along $x$ is given by [1]

$$R_{xy}^{2\omega} = R_{\text{DL+ANE}}^{2\omega}\cos(\varphi+A) + R_{\text{FL+Oe}}^{2\omega}\cos(2\varphi+2A)\cos(\varphi+A) + R_{\text{PNE}}^{2\omega}\sin(2\varphi+2A) + R_0$$

(S1)

with

$$R_{\text{DL+ANE}}^{2\omega} = -\frac{R_{\text{AHE}}H_{\text{DL}}^{\text{eff}}}{2(H_k+H_{\text{ext}})} - R_{\text{ANE}}^{2\omega} \quad \text{(S2)}$$

$$R_{\text{FL+Oe}}^{2\omega} = \frac{R_{\text{PHE}}H_{\text{FL+Oe}}^{\text{eff}}}{H_{\text{ext}}} \quad \text{(S3)}$$

where $R_{\text{AHE}}$ and $R_{\text{PHE}}$ denotes the anomalous Hall and planar Hall resistance of the sample, respectively; $H_k$ and $H_{\text{ext}}$ represent the anisotropy field and the applied magnetic field, respectively; $H_{\text{DL}}^{\text{eff}}$ denotes the effective field of current-induced dampinglike torques, while $H_{\text{FL+Oe}}^{\text{eff}}$ denotes the combined effective field of current-induced fieldlike torques and transverse Oersted fields; $R_{\text{DL+ANE}}^{2\omega}$ denotes the contributions to $R_{xy}^{2\omega}$ from the dampinglike torques and the anomalous Nernst effect $R_{\text{ANE}}^{2\omega}$; $R_{\text{FL+Oe}}^{2\omega}$ represent the contributions to $R_{xy}^{2\omega}$ from the fieldlike torques and the current-induced Oersted field; $R_{\text{PNE}}^{2\omega}$ is the contribution to $R_{xy}^{2\omega}$ from the planar Nernst effect; $A$ and $R_0$ denote the constant offset of $\varphi$ and $R_{xy}^{2\omega}$, respectively. Consequently, $H_{\text{DL}}^{\text{eff}}$ and $H_{\text{FL+Oe}}^{\text{eff}}$ can be extracted through linear fittings of $R_{\text{DL+ANE}}^{2\omega}$ and $R_{\text{FL+Oe}}^{2\omega}$ according to Eqns. (S2, S3), respectively. The dampinglike torque efficiency per unit current density $\xi_{\text{DL}}^{j}$ can be calculated as [2]

$$\xi_{DL}^{j} = \frac{2e}{\hbar} M_S t \frac{H_{DL, FL}^{eff}}{j} \quad (S4)$$

where $M_S$ and $t$ denotes the saturation magnetization and the thickness of the ferromagnetic layer, respectively. In our experiments, the $H_k$ of Cr/Tb at 150 K deduced from the anomalous Hall effect is ~8198 Oe, and we assume all the applied current flows in the Cr layer. Consequently, the $H_{DL}^{eff}$ and $\xi_{DL}^{j}$ are ~178.3 Oe and ~3.66 as described in the main text.

## 2 Fieldlike torques in Cr/Tb

The extracted $H_{FL+Oe}^{eff}$ of Cr/Tb is negligible as shown in Fig. S1. This indicates that orbital currents generated by the orbital Hall effect mainly give rise to dampinglike torques on magnetization, similar to the case of the spin Hall effect.

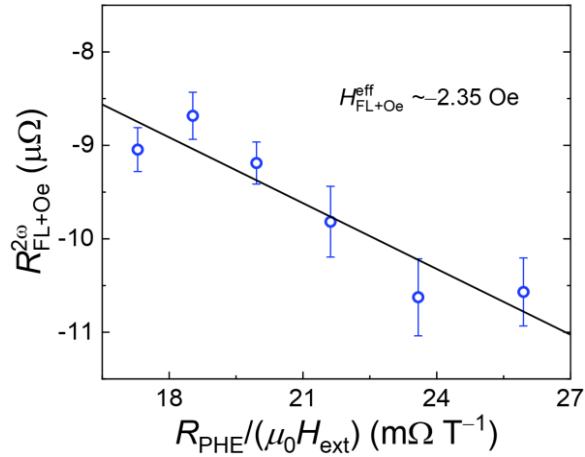

FIG S1. Linear fitting of $R_{FL+Oe}^{2\omega}$ with respect to $R_{PHE}/(\mu_0 H_{ext})$, which yields an effective field $H_{FL+Oe}^{eff}$ of ~–2.35 Oe. The meanings of the symbols are the same to those defined in section 1.

## 3 Dampinglike torques in Cr/Co

A control sample of MgO/Cr(10 nm)/Co(2 nm)/AlO$_x$(2 nm) was deposited to confirm the negligible effect of orbital currents on spin moments. As shown in Fig. S2, the $H_{DL}^{eff}$ in Cr/Co is subtle and opposite in sign with respect to that in Cr/Tb. The $\xi_{DL}^{j}$ is deduced to be ~–0.24, consistent with previous reports [3,4].

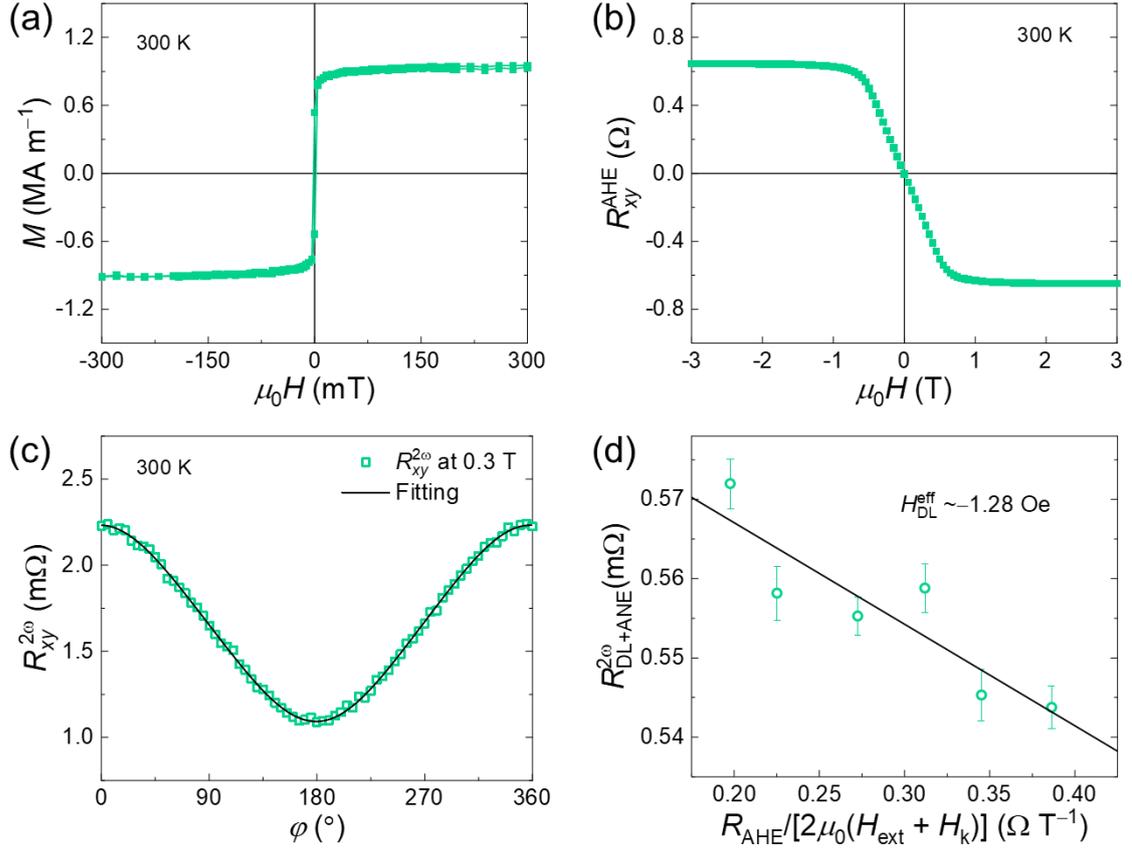

FIG S2. (a) In-plane magnetization $M$ and (b) Hall resistance after subtracting the ordinary Hall effect $R_{xy}^{\text{AHE}}$ as a function of applied magnetic fields $\mu_0 H$ of Cr/Co at room temperature. (c) Second-harmonic Hall resistance $R_{xy}^{2\omega}$ as a function of the angle $\varphi$ between the applied current $j \sim 3\times10^6$ A cm$^{-1}$ and $\mu_0 H$ at 300 K. The black lines are fitting curves according to Eqn. S1. (d) Linear fitting of $R_{\text{FL+Oe}}^{2\omega}$ with respect to $R_{\text{AHE}}/[2\mu_0(H_{\text{ext}} + H_k)]$, which yields an effective field $H_{\text{DL}}^{\text{eff}}$ of $\sim-1.28$ Oe. The meanings of the symbols are the same to those defined in section 1.